# Cyber Protection Applications of Quantum Computing: A Review

Ummar Ahmed, Tuomo Sipola, Jari Hautamäki

Institute of Information Technology, Jamk University of Applied Sciences, Jyväskylä, Finland

ummar.ahmed@student.jamk.fi
tuomo.sipola@jamk.fi
jari.hautamaki@jamk.fi

**Abstract:** Quantum computing is a cutting-edge field of information technology that harnesses the principles of quantum mechanics to perform computations. It has major implications for the cyber security industry. Existing cyber protection applications are working well, but there are still challenges and vulnerabilities in computer networks. Sometimes data and privacy are also compromised. These complications lead to research questions asking what kind of cyber protection applications of quantum computing are there and what potential methods or techniques can be used for cyber protection? These questions will reveal how much power quantum computing has and to what extent it can outperform the conventional computing systems. This scoping review was conducted by considering 815 papers. It showed the possibilities that can be achieved if quantum technologies are implemented in cyber environments. This scoping review discusses various domains such as algorithms and applications, bioinformatics, cloud and edge computing, the organization of complex systems, application areas focused on security and threats, and the broader quantum computing ecosystem. In each of these areas, there is significant scope for quantum computing to be implemented and to revolutionize the working environment. Numerous quantum computing applications for cyber protection and a number of techniques to protect our data and privacy were identified. The results are not limited to network security but also include data security. This paper also discusses societal aspects, e.g., the applications of quantum computing in the social sciences. This scoping review discusses how to enhance the efficiency and security of quantum computing in various cyber security domains. Additionally, it encourages the reader to think about what kind of techniques and methods can be deployed to secure the cyber world.

## 1. Introduction

Quantum technology is an emerging field that is changing the IT industry. The power of quantum technology can increase the computational capability of any computing environment and produce highly efficient results. Quantum computing is in a growth phase, but it still has enough power to impact real-world scenarios (Coccia, Roshani and Mosleh, 2024). One of these scenarios is software engineering. Quantum software engineering plays a pivotal role in building applications especially in the most complex and critical areas (Ali et al., 2022). Quantum computing also plays a critical role in concrete applications such as smart city planning and urban design. This is a less technical aspect of the world where quantum computing can deliver extraordinary results (Bashirpour Bonab et al., 2023a). Furthermore, quantum computing has taken problem solving to the next level. For example, Dempster–Shafer Theory (DST) shows that quantum computing can be used to analyse data using quantum mechanical principles (Zhou, Tian and Deng, 2023). Raheman (2022) describes the future of quantum computing in relation to cyber security. He highlights the concepts of post quantum cryptography, exploits against quantum-as-a-service and mitigating network architecture. As the cyber world expands, devices mature, and new technologies emerge, problems become more complex. (Shaw and Dutta, 2023). The question that arises is how we can use quantum computing in cyber protection. As every device on the network generates data, the likelihood of network vulnerabilities increases with the advancement of tools and skills. This review has been conducted precisely to explore the applications of quantum computing in the field of cyber protection, given the increasing challenges posed by the evolving cyber landscape.

The journey began with research questions aimed at understanding the cyber protection applications of quantum computing and explore potential methods or techniques for effective solutions in this domain. In this scoping review, databases such as IEEE Xplore, Science Direct, and Google Scholar were used to gather relevant information. Machine learning and artificial intelligence are key areas and the synergy with quantum computing opens up possibilities that promise unprecedented levels of excellence (Maheshwari, Garcia-Zapirain and Sierra-Sosa, 2022). This study uncovers the potential of quantum computing to address complex problem-solving scenarios and explores how quantum engineers can receive automated assistance in developing diverse quantum programs for problem-solving applications (Alonso, Sánchez and Sánchez-Rubio,

Funded by the European Union

2022). With networks constantly generating data and ubiquitous use of the Internet of things (IoT) in which numerous devices communicate with each other, the challenge of maintaining data protection becomes pronounced. This study focuses on into identifying quantum algorithms and techniques that can enhance security, mitigating risks and vulnerabilities in such interconnected environments (Li et al., 2022). Beyond the technical aspects, this study sheds light on the deceptive practices in the quantum realm. It draws attention to instances where certain companies are engaged in futile endeavours, highlighting the importance of distinguishing genuine advancements from misleading activities driven by financial motives (Khan et al., 2023). Throughout the scoping review, the following research questions will be addressed:

- **RQ1:** What are the applications of quantum computing in the field of cybersecurity, in particular cyber protection?
- **RQ2:** What are the potential methods or techniques that could be used to effectively address the challenge of cyber protection?

From here, our paper will follow to the methodology in section 2, which includes steps such as providing the study background, formulating the research question, and detailing the processes of data collection, identification, screening, and paper inclusion. Moving on to section 3, the results of this scoping review will be discovered, organized into different categories within its subcategories. Following this, section 4 deals with the discussion, while section 5 serves as the conclusion, summarizing all the findings of this scoping review.

## 2. Methodology

This is systematic scoping review conducted to examine the cyber protection applications of quantum computing. The stages of scoping review described by Arksey and O'Malley (Arksey and O'Malley, 2005; Levac et al., 2010) were used in this study.

Having research questions about the applications of quantum computing for cyber protection and the techniques for potential solutions, and recognizing the breadth of the topic, the exploration by employing targeted search phrases was initiated. IEEE Xplore, Science Direct and Google scholar were used and *Quantum Computing Applications* and *Quantum computing use cases* search phrases were selected to start of scoping review, as described in Figure 1. This search was most relevant to this study, and it was finalised on the basis of outcomes found with it in different databases. Moreover, the research was limited to journals only to obtain the most precise data. As quantum computing is one of the novel computing technologies, the search was limited to the years 2022 and 2023. Moreover, it was decided to limit the research to 10 pages for Google Scholar and the first 400 relevant results using Science Direct. In the identification phase, the criteria for filtering the papers was the relevance of the data provided to quantum computing.

Thus, any the search results that included quantum studies with principals from physics and other domains were easily excluded using these criteria. In addition to this physics material, collaboration of quantum computing was found in industries of *data science*, *artificial intelligence*, *algorithms and applications*, *bioinformatics*, *cloud and edge computing*, *complex system optimisation*, *cryptography and security*, *education and learning*, *engineering and automated process control*, *environment science*, *ethics and societal impact*, *finance and economics*, *fundamentals and reviews*, *hardware and architectures*, *healthcare*, *IoT*, *machine learning*, *material science*, *nanotechnology*, *networks*, *optimization problems*, *power system applications*, *renewable energy*, *robotics*, *simulations*, *smart cities and urban planning*, *space technology*, *specific industries* and *theoretical physics*. All these are well-known domains but the research was limited and continued with *software and programming*, *smart cities and urban planning*, and *IoT (internet of Things)*. All these categories with their references are presented in Table 1.

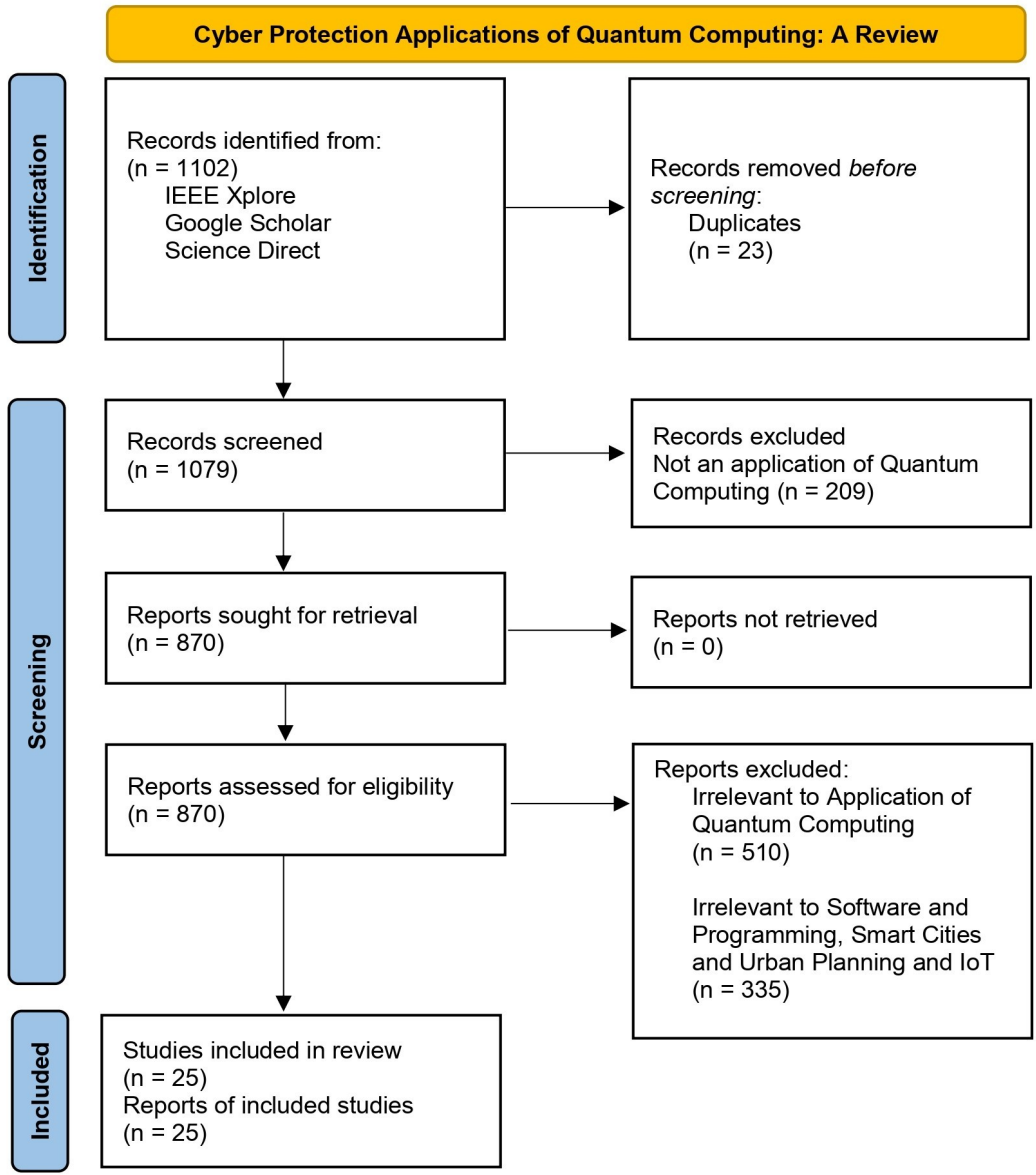

**Figure 1:** PRISMA flowchart of the scoping review

## 3. Results

The result for cyber protection applications of quantum computing covers many dimensions of emerging computing. It includes different categories such as data security, innovation in IT industry, energy, efficient use of energy, simulations, network traffic management and data security, quantum collaboration with emerging technology and most interestingly quantum application in social sciences. The resulting categorization is described in Table 1.

**Table 1:** Research Papers

| Categories | References |
|---|---|
| Generation of Quantum Programs | Alonso, Sánchez and Sánchez-Rubio (2022) |
| Data Security and Privacy | Shaw and Dutta (2023) |
|  | Li et al. (2022) |
|  | Padmaa et al. (2022) |
|  | Chawla and Mehra (2023) |
| Quantum Support Data | De Stefano et al. (2022) |
| Fake Improvements in Quantum Technology by IT giants | Khan et al. (2023) |
| Innovation in Computing Efficiency | Macrae (2023) |
|  | Willsch et al. (2022) |
|  | Zhou, Tian and Deng (2023) |
|  | Jałowiecki, Lewandowska and Pawela (2023) |
| Energy Efficiency | Al-Khafaji et al. (2023) |
| Problem Solving | Feng and Li (2023) |
|  | Pérez-Castillo, Jiménez-Navajas and Piattini (2023) |
|  | Yi et al. (2023) |
| Simulation | Arufe et al. (2023) |
| Network Communication | Hildebrand et al. (2023) |
|  | Qadir et al. (2022) |
|  | Abd El-Aziz et al. (2022) |
| Data Analytics | Harikrishnakumar and Nannapaneni (2023) |
| Quantum and Social Science | Bashirpour Bonab et al. (2023a) |
|  | Ukpabi et al. (2023) |
| Quantum and Emerging Technologies | Bashirpour Bonab et al. (2023b) |

### 3.1 Generation of Quantum Programs

Using quantum technology with Model Driven Engineering (MDE) techniques, such applications are developed that generate quantum programs. These programs are readable by quantum machines, and highly focused problems are proposed to be solved using this technique. Moreover, a metamodel has been introduced for quantum circuits and model-to-text transformations for generating IBM Qiskit code have been introduced, as they bring the quantum usability closer to reality (Alonso, Sánchez and Sánchez-Rubio, 2022).

### 3.2 Data Security and Privacy

Quantum continues to play its role for data security and privacy. Research proposed a quantum triple qubit model for image scaling using hue, saturation and light which encodes data in the scaled-up image and decodes data by scaling down at the receiver end (Li et al., 2023). Another research shows that as the types of interconnected devices increased, the chances of network attacks also increased due to the difficulty to manage security. Researchers called it signcryption which involves signing and encryption. This method uses isogeny-based encryption (Shaw and Dutta, 2023). Researchers proposed a new way to secure the user information in the advanced IoT environment using a technique called quantum key pool. The addition of advanced devices in networks motivated the researchers. This technique involves rapid distribution of encryption keys using quantum technology (Li et al., 2022)]. With the expansion of IoT networks, security and privacy are also compromised. Therefore, researchers come up with a new energy-efficient and secure data transmission protocol for intelligent IoT edge systems, called EEC-SDTP, to solve these challenges. This system will create cluster heads and routes for secure transmission. The system uses the oppositional chaos game optimization-based clustering (OCGOC) technique to determine cluster heads and routes (Padmaa et al., 2022) (Chawla and Mehra, 2023).

### 3.3 Quantum Support Data

A quantum software developer in today's world does not rely solely reliant on their own knowledge, but rather collaborates with a diverse community of researchers, engineers, and experts to advance the field. Quantum software development has such repositories and resources that can benefit the quantum developer in developing something e.g. ryuNagai/QML, oliverfunk/quantum-natural-gradient, kongju/QML etc. A study has been done to know how much quantum data is present in repositories to support quantum software engineers. Therefore, it is a positive gesture that quantum engineers share their work to support the quantum engineer community (De Stefano et al., 2022).

### 3.4 Fake Improvements in Quantum Technology by IT giants

At this time when quantum systems are being improved IT giants such as IBM, Google, or Microsoft are busy in exploiting the architectural solutions by developing various software to manage and manipulate quantum hardware. Quantum technology still has too much room for improvement. So-called IT giants are not really serving the industry. Most of the research papers published in the last 4 years are about solutions and validations. It is time to serve the quantum industry to get better results. (Khan et al., 2023).

### 3.5 Innovation in Computing Efficiency

Still, the quantum industry is growing in innovation. Quantum scientists are trying to build super-small and efficient computer systems using the quantum-dot cellular automata approach (Macrae, 2023). In addition, there is ongoing research to improve the computational power of computers. A new model called "balanced ternary" has been proposed to calculate the processing in a new way which will speed up the computational processing in an exceptional way (Faghih et al., 2023). In a GPU level research, researcher used Jülich universal quantum computer simulator (JUQCS–G) that was fast due to advanced GPUs. They tested the computer on JUWELS Booster, a GPU cluster with 3744 NVIDIA A100 Tensor Core GPUs and JUQCS-G to analyse the relationship between quantum annealing and quantum optimization algorithm. They surprisingly found approximate quantum annealing (AQA) (Willsch et al., 2022). Researchers try to explore the Dempster–Shafer Theory (DST) of AI, which is easy to represent knowledge but less useful in the real world due to a lot of data. To address this aspect, researchers use Basic Belief Assignment BBA and encode it in a quantum form which is derived from principles of quantum mechanics. It gave excellent results even for representing belief functions, analysing similarities between various information, collecting and evidence and transforming probabilities. It all happened because of quantum computing (Zhou, Tian and Deng, 2023). A Python library, PyQBench was introduced, which provides a ready-made command line interface CLI to use predefined functionalities. It can benchmark NICQ devices based on their capability of differentiating two von Neumann measurements (Jałowiecki, Lewandowska and Pawela, 2023).

### 3.6 Energy Efficiency

Research reveals an alternative called quantum dot cellular automata (QCA) to address traditional computing challenges. Researchers have successfully developed small energy-efficient digital circuits using QCA technology for IoT devices (Al-Khafaji et al., 2023).

### 3.7 Problem Solving

Quantum also plays an important role in problem solving. Research talks about error or problem checking techniques without actually running them. Researchers discuss "quantum-while-program" to show how these techniques are relevant to each other (Feng and Li, 2023). In addition, research also shows the quantum programs written in Python and D-Wave to analyse which problems are being solved by quantum annealing (Pérez-Castillo, Jiménez-Navajas and Piattini, 2023). Companies provide services and products. If someone does not plan when and where to use these services/products, they are useless. Researchers are trying to solve this problem with a technique called quantum annealing. Successful experiments have been carried out on the larger problems of vehicles (Yi et al., 2023).

### 3.8 Simulation

In discussions surrounding simulation, researchers propose a solution known as DBGA-X to address the quantum circuit compilation problem (QCCP), facilitating the execution of quantum programs on actual quantum hardware (Arufe et al., 2023).

## 3.9 Network Communication

Research has brought blockchain-enabled IoV (Internet of Vehicles) technology to overcome security and privacy issues on the Internet of Vehicles by not relying on a central entity. Vehicles communicate with each other to understand the traffic environment without sharing sensitive information (Hildebrand et al., 2023). The idea of 6G emerged to solve tricky problems in optimizing the network by using artificial intelligence, THz (terahertz) and quantum communications. As a result of this effort, holographic beamforming, AI-powered IoT networks, edge computing, and backscatter communications are expected (Qadir et al., 2022). In the world of neural networks, researchers are bringing a new technique called deep residual learning based quantum classical neural network (Res-QCNN). It will be used to monitor data coming from various devices and understand the IoT data in a better and faster way (Abd El-Aziz et al., 2022).

## 3.10 Data Analytics

Quantum is also used for prediction. One study shows how to optimise prediction for bike-sharing. Researchers used quantum Bayesian networks (a fancy term for a quantum computing method) to make the prediction more accurate over different days (Harikrishnakumar and Nannapaneni, 2023).

## 3.11 Quantum and Social Science

The approach known as the quantum urban paradigm seeks to tackle issues across multiple domains of liveability such as society, environment, and technology. By combining quantum technologies with social sciences, it aims to improve daily life by, for instance, optimising urban planning and streamlining communication management (Bashirpour Bonab et al., 2023a). While everyone is talking about the technical side of quantum computing, this paper adds a new dimension. It discusses the impact of quantum computing on business, politics, solving legal problems and other such non-technical areas (Ukpabi et al., 2023).

## 3.12 Quantum and Emerging Technologies

Researchers point out that quantum technologies can improve smart cities in terms of IoT, cloud computing, big data, smart transport, AI, and blockchain (Bashirpour Bonab et al., 2023b). Traditional von Neumann architectures can be used to integrate quantum technologies into modern smart city solutions. This integration of physics-based quantum technologies will represent a new paradigm that differs from current standards.

## 4. Discussion

Researchers have proposed a new way to keep the user information secure in an advanced IoT environment using the technique called quantum key pool. The Researchers felt the need for it due to the arrival of advanced devices in the network. This technique involves the rapid distribution of encryption keys using quantum technology. It is also a new way to keep smart grid data private on mobile devices. It uses a special method to manage secret keys and allows users to securely check their data securely. The proposed method is said to be fast, lightweight, and strong, using three layers of security. However, more research is needed to ensure that one part of the method works well (Li et al., 2022).

If we look at the security of network communications, researchers have introduced blockchain enabled internet of Vehicles technology to overcome security and privacy issues in the Internet of Vehicles by not relying on a central entity. Vehicles communicate with each other to understand the traffic environment without sharing sensitive information. In addition, the research explores recent advances in Blockchain-based Internet of Vehicles (BIoV) networks, covering applications like crowdsourcing, energy trading, and traffic management. It explores collaborative learning, blockchain-enabled hardware security, and the intersection of blockchain and quantum computing (Hildebrand et al., 2023). Researchers say that the evolution of wireless networks, such as 5G, is improving the technology, but it is not enough for the growing communication needs. Researchers are now looking at 6G networks to meet future demands, exploring technologies such as THz communication and quantum communication for faster data rates (Qadir et al., 2022). Other research has created a quantum-conventional hybrid neural network has been created, called Res-QCNN, to improve cost function optimization for deep networks. By using quantum neural networks in a residual structure block, the Res-QCNN training algorithm allows information to flow efficiently through layers, similar to deep residual learning in classical artificial neural networks, although it can be run on a standard computer (Abd El-Aziz et al., 2022).

Quantum computing is also making an impact on data security and privacy. In one study, researchers introduced a quantum colour model called quantum hue, saturation, and lightness (QHTS) based on triple-qubit sequence encoding. They developed quantum circuits for encoding and retrieving QHTS images using only a few qubits, improving operability. The study explored applications such as quantum steganography for secure message embedding and a spatial image fusion algorithm for remote sensing, demonstrating improved embedding capacity and potential advances in quantum image processing. Future work aims to explore quantum bicubic interpolation, improve the efficiency of key image generation, and enhance the quality of image fusion using advanced technologies. The researchers also plan to establish a quantum image fidelity metric (QIFM) for evaluating QHTS images in various applications (Li et al., 2023). Another study discusses the multi-user level encryption. In this study, the team has developed a quantum-resistant signcryption scheme that ensures key invisibility and ciphertext anonymity, addressing security concerns not addressed by many existing schemes. The size of their encrypted data is small, comparable to SeaSign signatures, and their security analysis accounts for insider attacks, multi-user settings, and non-repudiation. Additionally, key privacy has been investigated in an isogeny-based encryption scheme called Hashed-PKE (Shaw and Dutta, 2023).

During this scoping review, we also found a new way to secure data in smart grids using a method called PMAS-QS. It involves a multi-level privacy-preserving encryption, a quantum key pool for key management, and a data query method using proxy re-encryption. The proposed scheme is analysed to be efficient, lightweight, and robust, although the efficiency of the aggregate signcryption aspect is identified for further investigation (Li et al., 2022). In another research, a new technique called EEC-SDTP is introduced to make IoT edge systems more energy-efficient and ensures secure data transmission. The technique focuses on selecting optimal communication hubs (CHs) and secure routes for data transmission by using the OCGOC technique for CH selection and the SRP-QSPO technique for determining trustworthy routes. The inclusion of trust factors increases the reliability of node selection, resulting in superior performance compared to existing methods, as demonstrated by detailed simulation analyses. Future extensions could explore resource allocation and task scheduling strategies for more efficient use of available resources in IoT edge systems (Padmaa et al., 2022).

## 5. Conclusion

The research was started to get the answers to our research questions, the first one being that what are the cyber protection applications of quantum computing (RQ1). This is an important topic to study now, because of the advancement of network traffic and the versatility of data generation and communication of devices in the Internet (Internet of Things, IoT). There should be solutions or techniques to overcome these vulnerabilities. Therefore, we decided to ask that what kind of techniques or methods can be used for cyber protection (RQ2). Nowadays, different kinds of devices are connected to networks and produce data continuously. We have a fast, lightweight, and robust method with three layers of security called a quantum key pool for the security of user information. There are also links between blockchain and quantum technologies. In addition, cost function optimisation of deep networks has been improved using a quantum-conventional hybrid neural network. Light-based technologies for data security were presented, with applications in quantum steganography and spatial image fusion. A quantum-resistant signcryption scheme was also presented. In smart grids, this technology ensures secure data management. It will improve energy efficiency and secure data transmission in IoT edge systems with optimised communication hubs and routes, offering potential for further exploration in resource allocation strategies. The main limitation of this study is that it focuses only on journal articles. In technical areas, such as quantum computing, many important studies are also published in conferences. It would be beneficial to extend our research into different areas of quantum computing, including the exploration of quantum-safe standards, conducting security analyses using quantum computing, and investigating the application of blockchain in the field of quantum computing. Additionally, further studies could be conducted on the role of quantum computing in education and training.

**Acknowledgements**

We are grateful to Emils Bagirovs, Grigory Provodin and Thien Nguyen for their help with data collection and assistance with database creation. This research was partially supported by the ResilMesh project, funded by the European Union's Horizon Europe Framework Programme (HORIZON) under grant agreement 101119681. The authors would like to thank Ms. Tuula Kotikoski for proofreading the manuscript.

**References**

Abd El-Aziz, R.M., Taloba, A.I. and Alghamdi, F.A. (2022). "Quantum Computing Optimization Technique for IoT Platform using Modified Deep Residual Approach". *Alexandria Engineering Journal*, Vol 61(12), pp 12497–12509. https://doi.org/10.1016/j.aej.2022.06.029

Ali, S., Yue, T. and Abreu, R. (2022). "When software engineering meets quantum computing". *Communications of the ACM*, Vol 65(4), pp 84–88. https://doi.org/10.1145/3512340

Al-Khafaji, H.M.R. et al. (2023). "Performance optimization of the nano-scale carry-skip adder based on quantum dots and its application in the upcoming Internet of Things". *Optik*, Vol 287, pp 170976. https://doi.org/10.1016/j.ijleo.2023.170976

Alonso, D., Sánchez, P. and Sánchez-Rubio, F. (2022). "Engineering the development of quantum programs: Application to the Boolean satisfiability problem". *Advances in Engineering Software*, Vol 173, pp 103216. https://doi.org/10.1016/j.advengsoft.2022.103216

Arufe, L. et al. (2023). "New coding scheme to compile circuits for Quantum Approximate Optimization Algorithm by genetic evolution". *Applied Soft Computing*, Vol 144, pp 110456. https://doi.org/10.1016/j.asoc.2023.110456

Bashirpour Bonab, A. et al. (2023a). "In complexity we trust: A systematic literature review of urban quantum technologies". *Technological Forecasting and Social Change*, Vol 194, pp 122642. https://doi.org/10.1016/j.techfore.2023.122642

Bashirpour Bonab, A. et al. (2023b). "Urban quantum leap: A comprehensive review and analysis of quantum technologies for smart cities". *Cities*, Vol 140, pp 104459–104459. https://doi.org/10.1016/j.cities.2023.104459

Chawla, D. and Mehra, P.S. (2023). "A Survey on Quantum Computing for Internet of Things Security". *Procedia Computer Science*, Vol 218, pp 2191–2200. https://doi.org/10.1016/j.procs.2023.01.195

Levac, D., Colquhoun, H. and O'Brien, K.K. (2010). "Scoping studies: Advancing the Methodology". *Implementation Science*, Vol 5(1), pp.1–9. https://doi.org/10.1186/1748-5908-5-69

De Stefano, M. et al. (2022). "Software engineering for quantum programming: How far are we?" *Journal of Systems and Software*, Vol 190, pp 111326. https://doi.org/10.1016/j.jss.2022.111326

Faghih, E. et al (2023). "Efficient realization of quantum balanced ternary reversible multiplier building blocks: A great step towards sustainable computing". *Sustainable Computing: Informatics and Systems*, Vol 40, pp 100908. https://doi.org/10.1016/j.suscom.2023.100908

Feng, Y. and Li, S. (2023). "Abstract interpretation, Hoare logic, and incorrectness logic for quantum programs". *Information and Computation*, Vol 294, pp 105077. https://doi.org/10.1016/j.ic.2023.105077

Arksey, H. and O'Malley, L. (2005). "Scoping studies: Towards a Methodological Framework". *International Journal of Social Research Methodology*, Vol 8(1), pp.19–32. https://doi.org/10.1080/1364557032000119616

Harikrishnakumar, R. and Nannapaneni, S. (2023). "Forecasting Bike Sharing Demand Using Quantum Bayesian Network". *Expert Systems with Applications*, Vol 221, pp 119749. https://doi.org/10.1016/j.eswa.2023.119749

Hildebrand, B. et al. (2023). "A comprehensive review on blockchains for Internet of Vehicles: Challenges and directions". *Computer Science Review*, Vol 48, pp 100547. https://doi.org/10.1016/j.cosrev.2023.100547

Jałowiecki, K., Lewandowska, P. and Pawela, Ł. (2023). "PyQBench: A Python library for benchmarking gate-based quantum computers". *SoftwareX*, Vol 24, pp 101558. https://doi.org/10.1016/j.softx.2023.101558

Khan, A.A. et al. (2023). "Software architecture for quantum computing systems - a systematic review". *Journal of Systems and Software*, Vol 201, pp 111682. https://doi.org/10.1016/j.jss.2023.111682

Li, K. et al. (2022). "A novel privacy-preserving multi-level aggregate signcryption and query scheme for Smart Grid via mobile fog computing". *Journal of Information Security and Applications*, Vol 67, pp 103214. https://doi.org/10.1016/j.jisa.2022.103214

Li, N. et al. (2023). "Quantum image scaling with applications to image steganography and fusion". *Signal Processing: Image Communication*, Vol 117, pp 117015. https://doi.org/10.1016/j.image.2023.117015

Coccia, M., Roshani, S. and Mosleh, M. (2022). "Evolution of Quantum Computing: Theoretical and Innovation Management Implications for Emerging Quantum Industry". IEEE Transactions on Engineering Management, Vol. 71, pp.1–11. https://doi.org/10.1109/TEM.2022.3175633

Macrae, R.M. (2023). "Mixed-valence realizations of quantum dot cellular automata". *Journal of Physics and Chemistry of Solids*, Vol 177, pp 111303. https://doi.org/10.1016/j.jpcs.2023.111303

Maheshwari, D., Garcia-Zapirain, B. and Sierra-Sosa, D. (2022). "Quantum Machine Learning Applications in the Biomedical Domain: A Systematic Review". *IEEE Access*, Vol 10, pp 80463–80484. https://doi.org/10.1109/ACCESS.2022.3195044

Padmaa, M., Jayasankar, T., Venkatraman, S. et al (2022). "Oppositional chaos game optimization based clustering with trust based data transmission protocol for intelligent IoT edge systems". *Journal of Parallel and Distributed Computing*, Vol 164, pp 142–151. https://doi.org/10.1016/j.jpdc.2022.03.008

Pérez-Castillo, R., Jiménez-Navajas, L. and Piattini, M. (2023). "Dynamic analysis of quantum annealing programs". *Journal of Systems and Software*, Vol 201, pp 111683. https://doi.org/10.1016/j.jss.2023.111683

Qadir, Z. et al. (2022). "Towards 6G internet of things: Recent advances, use cases, and open challenges". *ICT Express*, Vol 9(3). https://doi.org/10.1016/j.icte.2022.06.006

Raheman, F. (2022). "The Future of Cybersecurity in the Age of Quantum Computers". *Future Internet*, Vol 14(11), pp 335. https://doi.org/10.3390/fi14110335

Shaw, S. and Dutta, R. (2023). "A quantum resistant multi-user signcryption scheme featuring key invisibility for Internet of Things". *Journal of Information Security and Applications*, Vol 76, pp 103549. https://doi.org/10.1016/j.jisa.2023.103549

Ukpabi, D. et al. (2023). "Framework for Understanding Quantum Computing Use Cases from A Multidisciplinary Perspective and Future Research Directions". *Futures*, Vol 154, pp 103277–103277. https://doi.org/10.1016/j.futures.2023.103277

Willsch, D. et al. (2022). "GPU-accelerated simulations of quantum annealing and the quantum approximate optimization algorithm". *Computer Physics Communications*, Vol 278, pp 108411. https://doi.org/10.1016/j.cpc.2022.108411

Yi, L. et al. (2023). "Service provision process scheduling using quantum annealing for technical product-service systems". *Procedia CIRP*, Vol 116, pp 330–335. https://doi.org/10.1016/j.procir.2023.02.056

Zhou, Q., Tian, G. and Deng, Y. (2023). "BF-QC: Belief functions on quantum circuits". *Expert Systems with Applications*, Vol 223, pp 119885. https://doi.org/10.1016/j.eswa.2023.119885